\documentclass[aps,pra,twocolumn,superscriptaddress]{revtex4}

\usepackage[normalem]{ulem}
\usepackage{amsmath,amssymb}
\usepackage{multirow}
\usepackage{xcolor,soul}
\usepackage{srcltx}
\usepackage{hyperref,graphicx}

\def\be{\begin{equation}}
\def\ee{\end{equation}}
\newcommand{\ket}[1]{|#1\rangle}

\newcommand{\fisicarm}{Dipartimento di Fisica, Sapienza Universit\`{a} di Roma, Piazzale Aldo Moro, 5, I-00185 Roma, Italy}
\newcommand{\fisicami}{Dipartimento di Fisica, Politecnico di Milano, Piazza Leonardo da Vinci, 32, I-20133 Milano, Italy}
\newcommand{\pisasns}{NEST, Scuola Normale Superiore and Istituto di Nanoscienze - CNR, I-56126 Pisa, Italy}
\newcommand{\ino}{Istituto Nazionale di Ottica, Consiglio Nazionale delle Ricerche (INO-CNR), Largo Enrico Fermi, 6, I-50125 Firenze, Italy}
\newcommand{\ifn}{Istituto di Fotonica e Nanotecnologie, Consiglio Nazionale delle Ricerche (IFN-CNR), Piazza Leonardo da Vinci, 32, I-20133 Milano, Italy}

\begin{document}

\title{Anderson localization of entangled photons in an integrated quantum walk}

\author{Andrea Crespi}
\affiliation{\ifn}
\affiliation{\fisicami}

\author{ Roberto Osellame}
\email{roberto.osellame@polimi.it}
\affiliation{\ifn}
\affiliation{\fisicami}

\author{ Roberta Ramponi}
\affiliation{\ifn}
\affiliation{\fisicami}

\author{Vittorio Giovannetti}
\affiliation{\pisasns}

\author{Rosario Fazio}
\affiliation{\pisasns}

\author{Linda Sansoni}
\affiliation{\fisicarm}

\author{Francesco De Nicola}
\affiliation{\fisicarm}

\author{Fabio Sciarrino}
\email{fabio.sciarrino@uniroma1.it}
\affiliation{\fisicarm}
\affiliation{\ino}

\author{Paolo Mataloni}
\affiliation{\fisicarm}
\affiliation{\ino}

\maketitle

\textbf{Waves fail to propagate in random media.  First predicted for quantum particles in the presence of a disordered potential, 
Anderson localization  has been observed also in classical acoustics, electromagnetism and optics.  Here, for the first time, we 
report the observation of { Anderson localization} of pairs of { entangled photons} in a two-particle {discrete} quantum walk 
affected by position dependent disorder.  A quantum walk  {on} a disordered lattice {is} realized by {an} integrated array 
of interferometers {fabricated} in glass by femtosecond laser writing. {A} novel technique {is used} to introduce a controlled phase 
shift into each unit mesh of the network. Polarization entanglement {is} exploited to simulate the different symmetries 
of the two-walker system. {We are thus able to experimentally investigate the genuine effect of (bosonic and 
fermionic) statistics in the absence of interaction between the particles.} We will show how  different types of randomness and the symmetry of the wave-function affect the 
localization of the entangled walkers. }

In 1958 P.W.  Anderson \cite{ande58pr} predicted that the {wave-function of} a quantum particle can be localized in the presence of a static disordered 
potential.  As a consequence  of this mechanism it is expected that particle and energy transport through a disordered medium should 
be  strongly suppressed and that an  initial{ly} localized wave packet should not   spread out with time. After more than fifty years from 
its discovery Anderson localization  is still widely studied and it has pervaded many different areas of physics ranging from condensed 
matter and cold atomic physics to wave dynamics and quantum chaos~\cite{Anderson-localization-book}. This phenomenon  emerges 
quite generically  in the behavior of waves  in complex media, 
\begin{figure}[t]
\includegraphics[width=85mm]{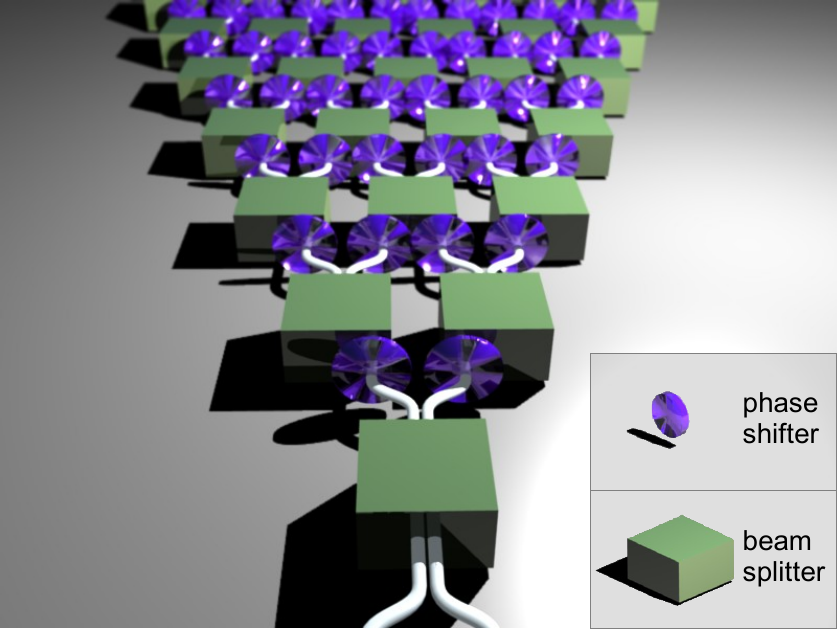}
\caption{ Concept scheme of cascaded beam splitters to implement photonic quantum walks. Disorder is introduced by phase shifters placed at each beam 
splitter's output port, before entering the next one.}
\label{fig:figure1}
\end{figure}
and it  has been experimentally observed in a variety of different systems: Bose-Einsten condensates~\cite{roat08nat,bill08nat}, 
light in semiconductor powders~\cite{wier97nat} and photonic lattices~\cite{pert04prl,schw07nat,lahi08prl}, single photons 
in bulk optics~\cite{broo10prl} and in fiber loops~\cite{schr11prl}, microwaves in strongly scattering samples~\cite{stor06prl}, besides 
ultrasound waves in a three-dimensional elastic system~\cite{hu08npy}. 

Anderson localization is a single-particle process which arises   from the destructive interference among different scattering paths.
Nevertheless, even in the absence of a direct interaction between particles,
pure quantum correlations~\cite{footnote} are  expected to influence in a non-trivial way the underlying  localization 
dynamics~\cite{omar06pra,been09prl,lahi10prl,brom09prl}. By taking advantage of the perfect phase stability provided by miniaturized 
integrated waveguide circuits~\cite{sans12prl}, we experimentally simulate a quantum walk of  a  two-photon {polarization-entangled} 
state  in a disordered medium. 
We are thus able, through a mapping derived in Ref.~\cite{omar06pra},  to 
test the localization of a pair of  non interacting particles obeying bosonic/fermionic statistics~\cite{shep94prl}.

\begin{figure*}[t!!]
\includegraphics[width=0.98\textwidth]{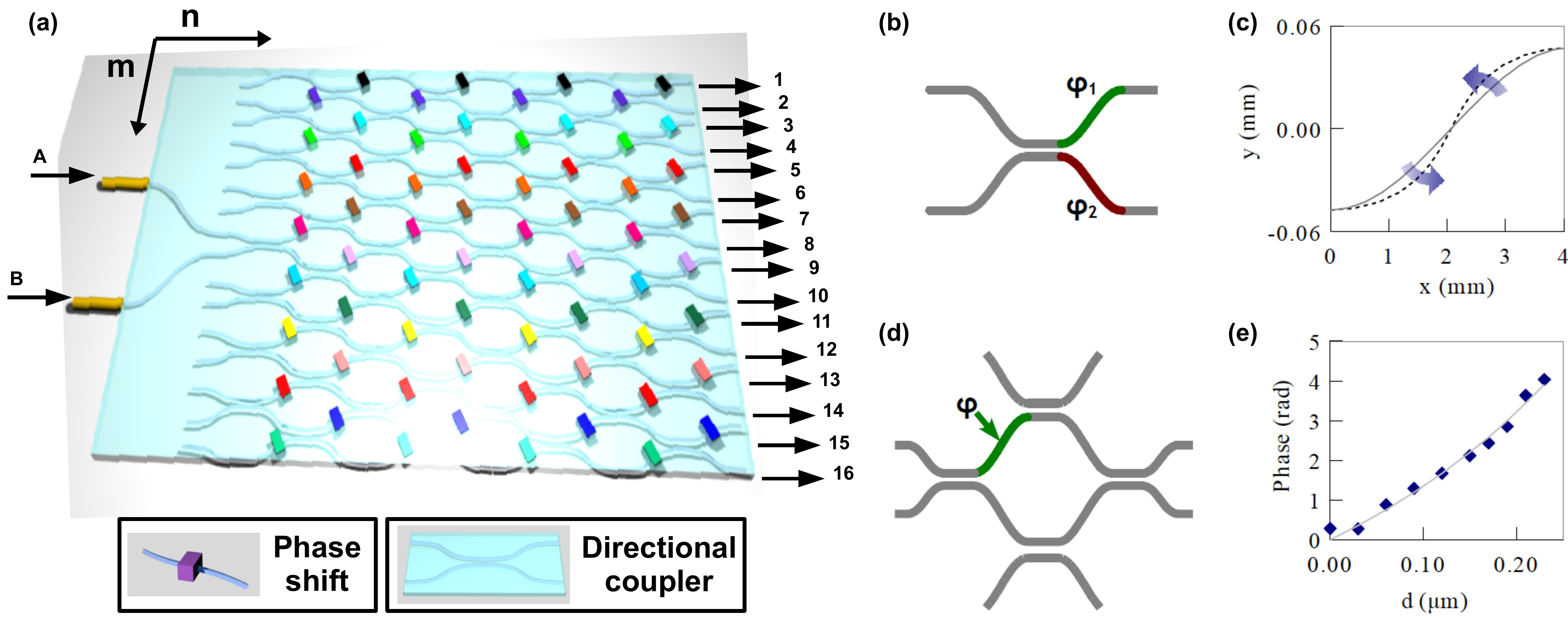}
\caption{{(a) Scheme of the network of directional couplers implementing a 8-step quantum walk with static disorder. Different colors stand for different phase shif{t}s. (b) Controlled deformation of either of the two S-bent waveguides at the output of each directional coupler extends the optical path and is equivalent to the application of a phase shifter. (c) The deformation is given by a non-linear coordinate transformation, which is function of a deformation coefficient $d$ (see Supplementary Information). The graph shows the undeformed S-bend (solid line), together with a deformed one (dashed). (d) Schematic of the Mach-Zehnder structure, representing the unit cell of the directional couplers network, fabricated for calibrating the phase shift induced by the deformation. (e) Phase shift induced by the deformation: theoretical curve calculated from the nominal geometric deformation (solid line), and experimental measurements (diamonds).}}
\label{fig:figure2}
\end{figure*}

A quantum walk (QW)~\cite{kemp03cph} is an extension of the classical random walk, where the walker goes back and forth along a line {and the} 
direction at each step depends on the result of a fair coin flip. At the quantum level, interference and superposition phenomena lead to a non-classical 
behavior of the walker giving  rise to new interesting effects, which can be harnessed to exponentially  speed up search 
algorithms~\cite{poto09pra} and to realize universal quantum computation~\cite{chil09prl}.  Besides, QWs have also been proposed to analyze 
energy transport in biological systems \cite{mohs08jcp,rebe09njp}.
Different experimental implementations of single-particle quantum
walks were performed with trapped atoms \cite{kars09sci}, ions \cite{schm09prl,zahr10prl}, energy
levels in NMR schemes \cite{ryan05pra}, photons in waveguide structures \cite{pere08prl}, in bulk optics \cite{broo10prl,kita12nco},
{and} in a fiber loop configuration \cite{schr10prl,schr11prl,schr12sci}. Very recently quantum walks of
two identical photons have been performed only {in} ordered structure{s}
\cite{sans12prl,peru10sci,owen11njp}.\\
A physical realization  of a discrete QW  {can be} provided by photons passing through a cascade of balanced  beam splitters {(BSs)}
arranged in a network of Mach-Zehnder (MZ) interferometers as shown in Figure \ref{fig:figure1}.
Here each BS implements simultaneously the quantum coin operation, i.e. the choice of the direction the particle will move in, and the step operator, which 
shifts the walker in the direction fixed by the quantum coin state (the time 
evolution being simulated stroboscopically) \cite{sans12prl}. 
Accordingly  every output of a BS of the network corresponds to a given point in the space-time of the QW, the horizontal {rows} of the setup representing different time steps. In this scenario, 
disorder  can be added in the  QW evolution
 by simply introducing (randomly selected)  phase shifts between the MZ interferometers  paths {(see Fig.~\ref{fig:figure1})}.  
{In particular the time-independent, \textit{static}, disorder needed to  enforce Anderson localization on the photonic  walker,  is obtained by {fixing} the 
same phases for all the MZs which {correspond to the same lattice site}.
Making sure that response of the device is polarization independent, 
the localization  of a non interacting entangled pair can now be studied by injecting in two different ports of the device -namely $A$ and $B$ in Fig \ref{fig:figure2}a- a two-photon entangled-polarized {state} generated via spontaneous parametric down-conversion.}
The above approach would be
extremely hard to implement with bulk optics mainly because of size
and of very challenging stability issues. However in the last few
years integrated quantum
photonics {proved} to be a highly promising experimental platform for
quantum information science \cite{obri09npo}. Recently, integrated waveguide
circuits  have been employed for quantum applications, in order to realize
two-qubit gates \cite{poli08sci,lain10apl,sans10prl,cres11nco}, quantum
algorithms \cite{poli09sci}, quantum walk on a chip \cite{peru10sci,matt11arx,sans12prl}
and enhanced quantum sensitivity in phase-controlled interferometers
\cite{smit09oex,matt09npo,cres12prl}. In order to observe Anderson localization for polarization entangled photons an important step forward is required in the 
available experimental platform. Our setup, for the first time,  integrate all the necessary ingredients to this aim:  polarization 
independent elements, interferometric structures, together with a proven capability to implement
suitable phase shift in different points of the QW circuit. Furthermore, in order to get a convincing evidence of localization it has 
been necessary to observe the experimental simulation for quantum walk of different steps. We were able to 
realize up to eight-step QW circuits affected by a controlled disorder, thus integrating tens of  BSs on the same chip.
 
In our experiment the setup of Fig{ure~\ref{fig:figure1}}
has been realized by using integrated waveguide circuits{, as in Figure~\ref{fig:figure2}a, where BS elements are replaced by directional couplers. The discrete $m$-axis indicates the different sites of the QW, while the discrete $n$-axis identifies the different time steps. The integrated waveguide circuits have been fabricated} by 
femtosecond laser writing technology \cite{gatt08npo,dval09joa}. {This technology exploits nonlinear absorption of femtosecond pulses, focused below the surface of a transparent dielectric substrate, to obtain a permanent and localized refractive index increase. Translation of the sample under the laser beam along the desired path enables the fabrication of optical waveguide circuits with arbitrary three-dimensional geometries.}
{I}n order to obtain a totally polarization independent behavior, 
 the 3-dimensional geometry detailed in Ref.~\cite{sans12prl} has been adopted {(}these devices 
are known to allow the propagation of polarization entangled states~\cite{sans10prl}).
The phase shifters are implemented by deforming {one} of the S-bent waveguides at the output of each directional coupler (green {or} red segments in Fig{ure} \ref{fig:figure2}b), in order to stretch the optical path. {The phase shift $[-\pi,\pi]$ in each MZ cell is implemented by lengthening the optical path in the green segment to introduce a $[0,\pi]$ phase shift, while the complementary range $[-\pi,0]$ is achieved by lengthening the red segment. In this way, smaller deformations, always of the same kind (lengthening of the path), are capable to provide the full range of phase shifts.}

\begin{figure*}[t!!!!]
\includegraphics[width=0.98\textwidth]{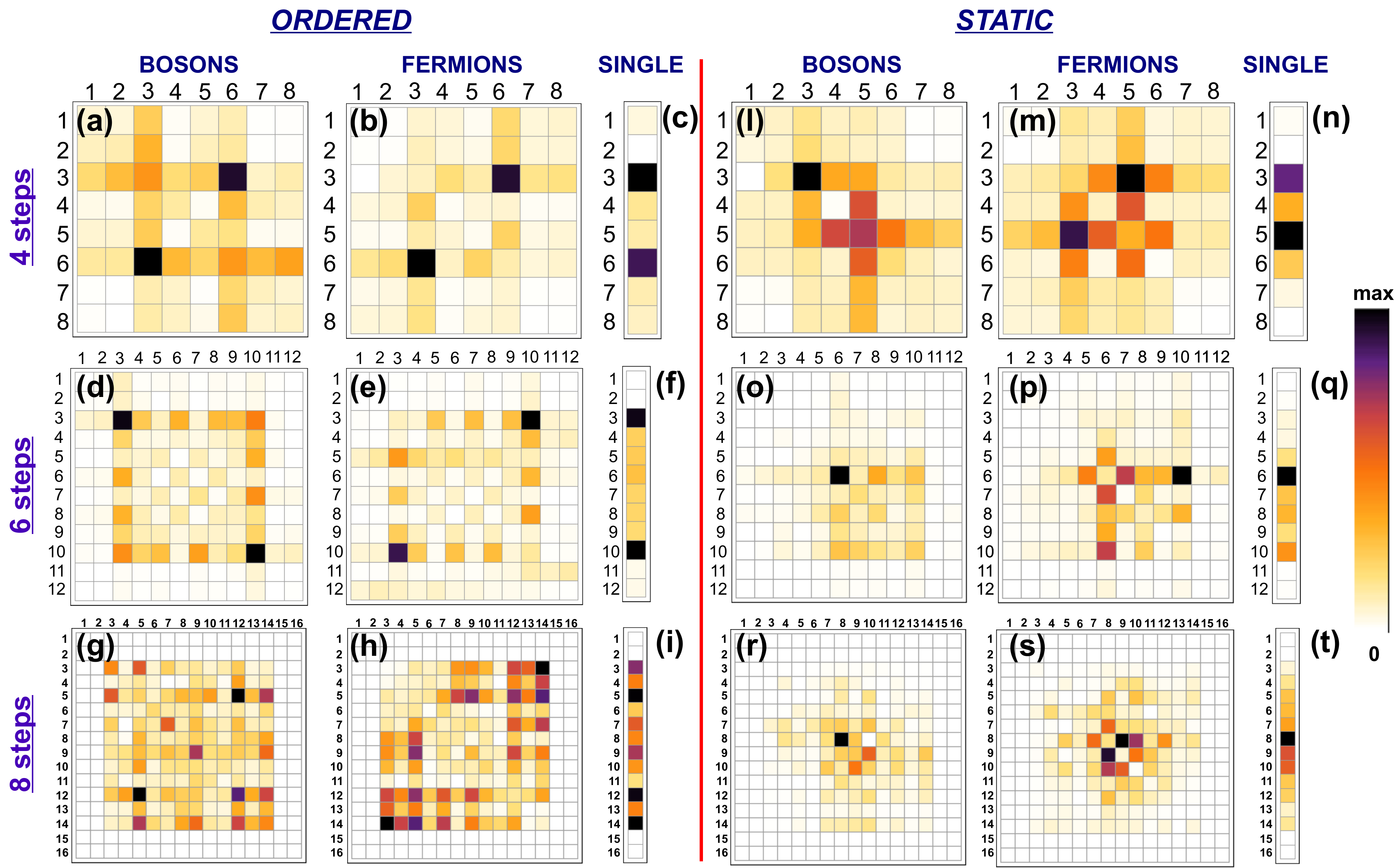}
\caption{Experimental results of single- and two-photon distributions for bosons and fermions in an ordered QW (a-i) and in a QW in presence of static disorder (l-t). Single-particle density distributions have been computed by tracing out the position of one of the particles (summing over the columns of the two-photon probability distribution matrices).}
\label{Figura3}
\end{figure*}
Figure \ref{fig:figure2}c shows both an undeformed and a deformed S-bend.
\begin{table}[h]
\begin{center}
\begin{tabular}{|c|}
\hline
\textbf{ORDERED}\\\hline
	\begin{tabular}{c|c|c}
	\it{Steps} & Input $A$ & Input $B$\\\hline
	4 & $0.991\pm0.002$ &  $0.992\pm0.002$\\\hline
	6 & $0.994\pm0.003$ &  $0.987\pm0.003$ \\\hline
	8 & $0.951\pm0.004$ &  $0.946\pm0.005$ \\
	\end{tabular}\\\hline
\textbf{STATIC}\\\hline
	\begin{tabular}{c|c|c}
	\it{Steps} & Input $A$ & Input $B$\\\hline
	4 & $0.980\pm0.003$ &  $0.976\pm0.002$ \\\hline
	6 & $0.985\pm0.002$ &  $0.976\pm0.003$ \\\hline
	8 & $0.938\pm0.004$ &  $0.957\pm0.004$ \\
	\end{tabular}\\\hline
\end{tabular}
\end{center}
\caption{Similarities for single particle QW distributions. The values are calculated as mean average on distributions of single-photon in different polarization states.}
\label{tab:single}
\end{table}

To test our technique and calibrate the achieved phase shift as a function of the imposed deformation $d$ {(see Supplementary Information for a detailed definition)} several MZ interferometers were fabricated with the design of Figure \ref{fig:figure2} d), reproducing exactly the unit cell of the QW network. Each interferometer has one S-bend (the one colored in the figure) deformed with a different value of $d$. Laser light at $\lambda=$806 nm wavelength was injected in the interferometers and the induced phase shift was then retrieved from the measured light distribution at the output.
Figure \ref{fig:figure2} e) reports the experimentally measured phase shifts as a function of the deformation parameter $d$. The experimental points are in good agreement with the phase shift predicted by evaluating numerically the geometric lengthening $\Delta l$ of the deformed S-bend $\phi_{theo} = \frac{2 \pi}{\lambda} \Delta l$.
\begin{table*}[htdp]
\begin{center}
\begin{tabular}{|c|c|}
\hline
\textbf{ORDERED} & \textbf{STATIC}\\\hline
\begin{tabular}{c|c|c|c}
	\it{Steps} & \it{Bosons} & \it{Fermions} & \it{Single}\\\hline
	4 & $0.946\pm0.003$ & $0.914\pm0.003$ & $0.996\pm0.001$\\\hline
	6 & $0.940\pm0.003$ & $0.851\pm0.003$ & $0.997\pm0.001$\\\hline
	8 & $0.768\pm0.006$ & $0.780\pm0.007$ & $0.934\pm0.004$\\
	\end{tabular}
	&
\begin{tabular}{c|c|c|c}
	\it{Steps} & \it{Bosons} & \it{Fermions} & \it{Single}\\\hline
	4 & $0.918\pm0.003$ & $0.902\pm0.003$ & $0.993\pm0.001$\\\hline
	6 & $0.890\pm0.006$ & $0.903\pm0.004$ & $0.985\pm0.002$\\\hline
	8 & $0.803\pm0.004$ & $0.785\pm0.004$ & $0.947\pm0.002$\\
	\end{tabular}\\\hline
\end{tabular}
\end{center}
\caption{Similarities between the experimental distributions of Fig. \ref{Figura3} and the expected ones. Uncertainties arise from the Poisson distribution of counting statistics.}
\label{tab:sim}
\end{table*}

\begin{figure}[h!!]
\includegraphics[width=\columnwidth]{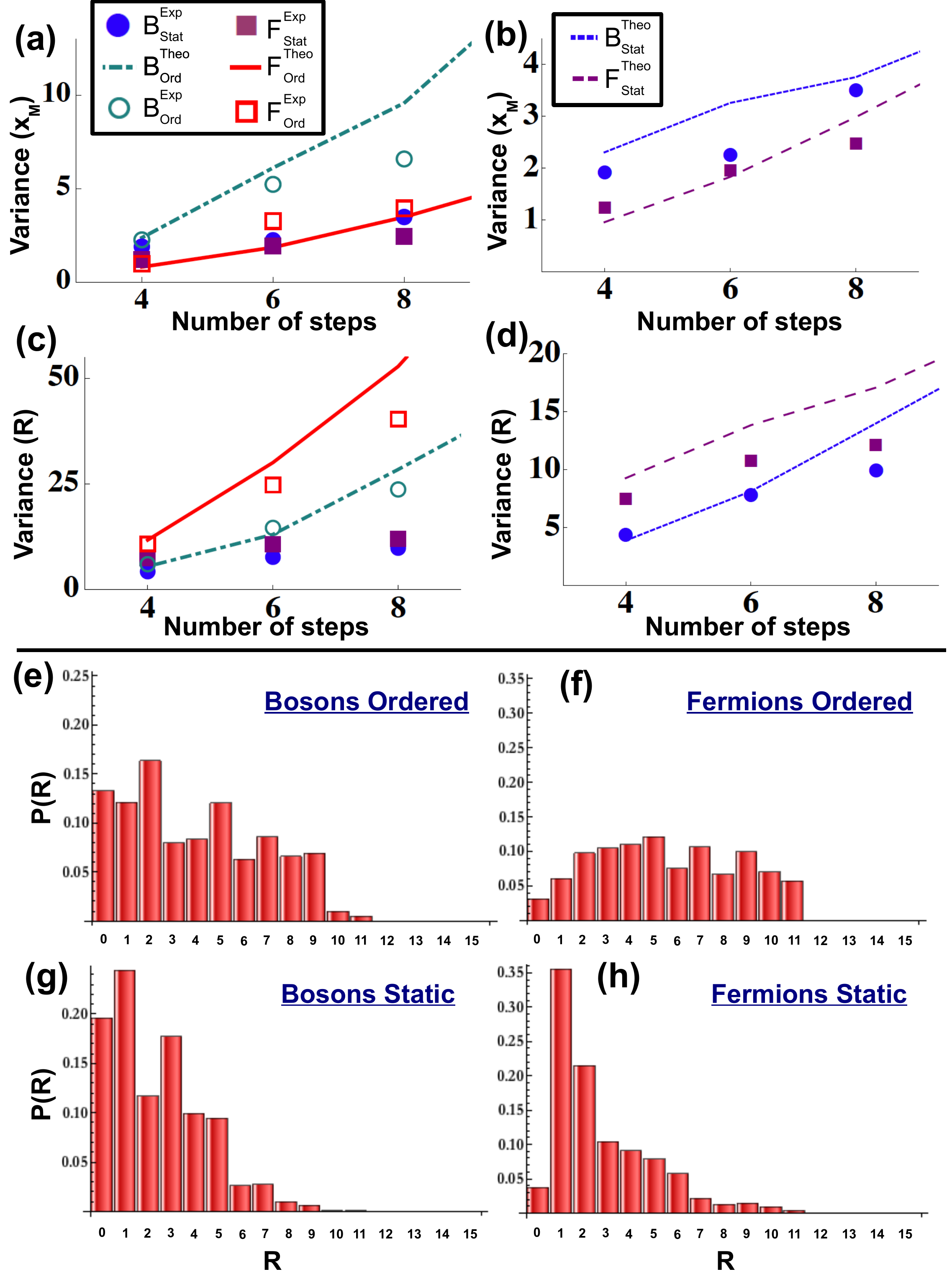}
\caption{
{(a-b) Variance of the two photon mean position $x_M=(j+k)/2$ and (c,d) of the relative distance $R=|i-k|$ shown in function of the number of steps. Experimental results are reported for bosons (circles) and fermions (squares). Empty markers refer to ordered QWs, whereas filled ones correspond to QWs with static disorder. In panels a,b dot-dashed and solid lines represent theoretical behaviors for bosons and fermions in the ordered case while in panels c,d dotted and dashed lines correspond to theory for bosons and fermions in presence of static disorder, respectively. Error bars are smaller than the spot size. (e-h) Probability distributions of the relative distance $R$ for bosonic (left) and fermionic (right) two-photon 8-step QWs in the case of a ordered QW (e-f) and in presence of position-dependent disorder (g-h).}}
\label{Figura4}
\end{figure}

We implemented a lattice with static disorder by imposing the same phase shift to the MZ cells corresponding to a fixed 
site of the QW line {as} in Fig. \ref{fig:figure2}a ($\phi_{m,n}=\phi_{m}, \forall n$). {QW circuits composed by 4, 6 and 8 step affected by static disorder were realized in a way that the 4-step phase pattern was embedded within the 6-step phase pattern and, in turn, this was embedded within the 8-step one.\\
Another set of 4, 6 and 8 step ordered, i.e. with 
perfectly symmetric MZ cells, QW circuits
was realized and compared with {the corresponding} disordered one.}

First of all we measured the single particle distributions (see Supplementary Information) in order to demonstrate the polarization insensitivity 
of the integrated QWs. We repeated this measurement by injecting single photon states with different polarizations. The measured distributions 
exhibit very similar behaviors.
{We compared the obtained results with the expected ones by calculating the similarity defined as $S=(\sum_{i,j}{\sqrt{D_{ij}D_{i,j}^{\prime}}})^2/\sum_{i,j}{D_{ij}}\sum_{i,j}{D_{ij}^{\prime}}$, which is a generalization of the  classical fidelity between two distributions $D$ and $D^{\prime}$.}
The obtained values are reported in Table \ref{tab:single} for the ordered QW 
circuits and for QWs with static disorder. These high values {and low deviations} highlight once more the {fabrication control and} polarization insensitivity of our integrated devices.

As a second step we carried out the investigation of two-particle QWs.
The investigation of  {Anderson localization} for bosonic and fermionic particles was realized by adopting the complete experimental 
setup described in details in the Supplementary Information.
As mentioned above, we exploited polarization entanglement to simulate bosonic and fermionic statistics.
To this {aim} 
{polarization-entangled} photon pairs{, generated via spontaneous parametric down-conversion,} were simultaneously injected into arms $A$ and $B$ of the {4, 6 and 8 steps} QW circuits to observe the progressive quench of 
photon propagation in disordered QWs {(Fig. \ref{fig:figure2}a)}. By setting 
the phase $\phi$ of the state $\frac{1}{\sqrt{2}}[\ket{H}_A\ket{V}_B+e^{i\phi}\ket{V}_A\ket{H}_B]$, bosonic ($\phi=0$) or fermionic ($\phi=\pi$) QWs were {observed}.\\
In Figure~\ref{Figura3} we show how entangled photons localize in the presence of a random static potential by plotting the joint probability $P_{j,k}$ of
detecting one particle in the output port $j$ and the other in the output port $k$ {(}the probability being determined by collecting events independently from the photon polarization{)}. 
The different panels 
compare the ordered and disordered cases in the case of symmetric and antisymmetric wave-functions. {W}e report also 
the case of single photons which {are} reconstructed by tracing out the position of one of the particles of the entangled pair. 
While in the case of an ordered 
system the walkers spread on with the increasing number of steps, Anderson localization implies that the wave-packets will remain localized around 
the central sites irrespectively of the number of steps. This is indeed what we observed. The difference between the ordered and disordered 
case is most evident for the 8-step QW, compare the panels (g) and (h) with (r) and (s) in  Figure~\ref{Figura3}. The agreement of the experimental 
data with the theoretical predictions, again quantified by the similarities, is reported in Table \ref{tab:sim}. {In the ordered case $\mathcal{S}$ is slightly
worse for the 8-step QW. This discrepancy, due to some unavoidable uncertainty in BSs realization is milder in the disordered 
case. Here, as expected, additional phase-shift to the ``intentionally-chosen" random one will {have less effect due to} localization.} 

\begin{figure}[ht!!!]
\includegraphics[width=0.95\columnwidth]{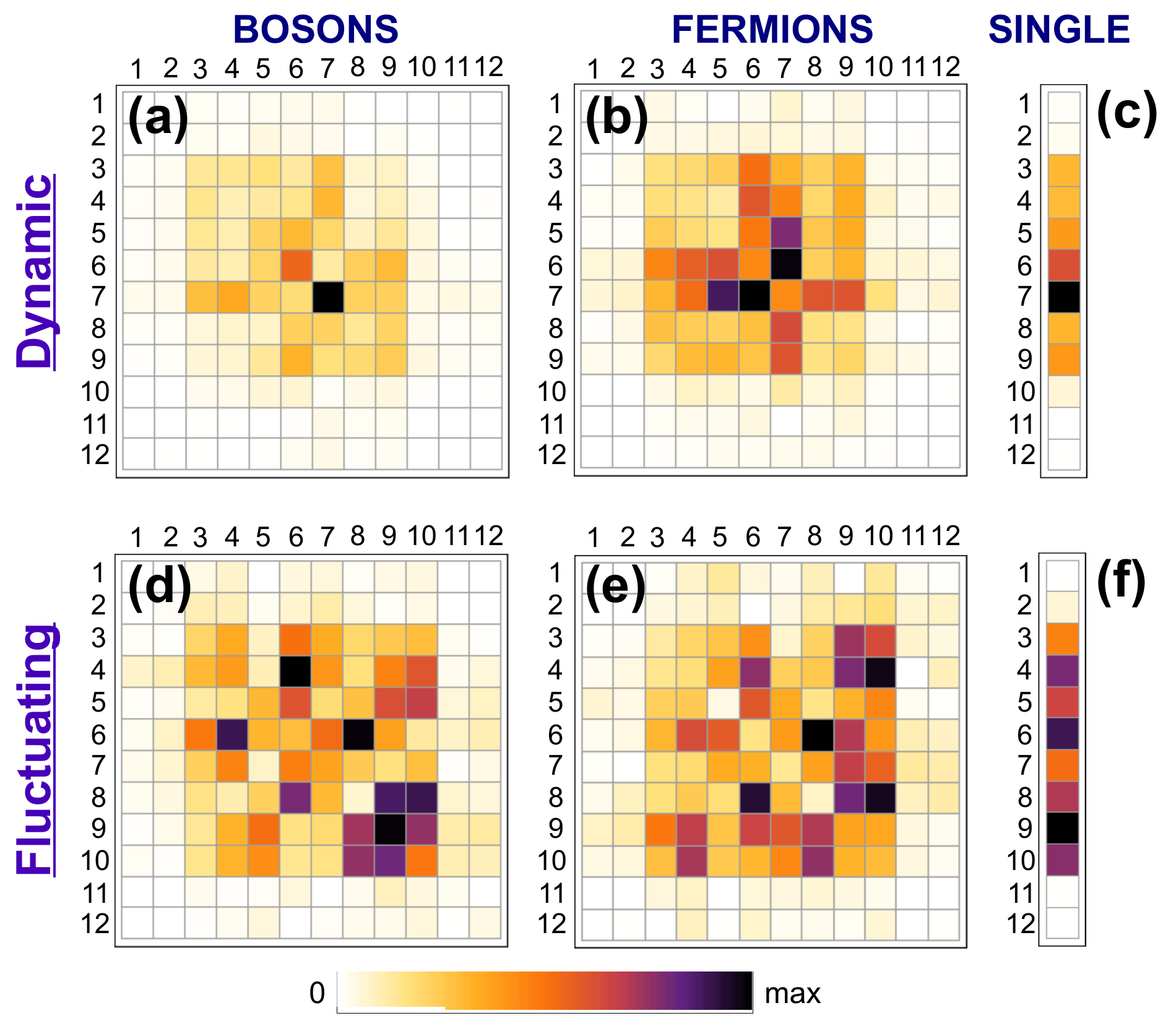}
\caption{Experimental results: single- and two-particle distributions for a 6-step QW in presence of dynamic (a-c) and fluctuating (d-f) disorder.
Single particle distributions have been computed by tracing out the position of one of the particles (summing over the columns of the probability distribution matrices) The similarities for bosons, fermions and single particle with the expected distributions are respectively $\mathcal{S}_{Bos}^D=0.871\pm0.004$,  $\mathcal{S}_{Fer}^D=0.802\pm0.006$ and  $\mathcal{S}_{Sing}^D=0.975\pm0.003$ for the QW circuit with dynamic disorder and  $\mathcal{S}_{Bos}^F=0.921\pm0.004$, $\mathcal{S}_{Fer}^F=0.852\pm0.003$ and $\mathcal{S}_{Sing}^F=0.991\pm0.002$ for the chip with fluctuating disorder.}
\label{Figura5}
\end{figure}

The entangled pairs localize in a manner which depends on their statistics.  
 A more quantitative estimate of the difference in the  localization properties of entangled photons may be obtained by looking at the
variance of the two-photon mean position $x_{M} = (j +k)/2$   associated with {the} probability distribution{s} of Figure~\ref{Figura3} as a function of the number of steps.  This is shown in Figure~\ref{Figura4}{a,b}. While 
for the ordered case the variance grows quadratically with the number of steps $n$, it is weakly dependent on $n$ in the disordered case, indicating that 
the system tends towards localization. The numerical simulation{,} performed by considering discrete time QWs {with static disorder} shows that localization starts even with a relatively small number of steps, giving rise to a fully localized state {(}corresponding to
a variance not varying with $n${)} typically for $n\sim 100$. However, such number of steps is currently out-of-reach for any technological platform. The clear difference observed in our experiment between ordered and disordered QWs is a strong evidence of the onset of Anderson localization for a small number of steps.\\
{In addition we observe} a very interesting 
feature emerging from the data, both in Fig. \ref{Figura3} and in Fig. \ref{Figura4}, {further evidencing} the different behaviour between bosons and fermions.
{By looking at the behavior of $x_M$, fermions localize more than bosons. Antisymmetry does help localization. This fact that may sound counterintuitive can be understood (see the Supplementary Information) by looking at the sign of the interference term on the variance of $x_M$.
An opposite behavior is observed for the distribution of the relative distance $R = |j-k|$ between the two particles. Because of the Pauli exclusion principle, the average distance between the particles in the fermionic case is larger than in the bosonic case (where they tend to bunch). This is exactly what we observe by looking at the variance of the distribution of the relative distances (Fig. \ref{Figura4}c,d), which follows from the probability distribution $P(R)$.
The different experimental distributions of the distance between the particles, obtained in both the ordered and static-disordered case, and the different behaviour of bosons and fermions are shown in Fig. \ref{Figura4}e-h.}\\
{So far we discussed the case of static disorder, however d}ifferent types of disorder affect differently the dynamics of the entangled pairs. {Since our technology is capable of implementing arbitrary phase maps in the QWs, a lattice with dynamic disorder was produced applying the same phase shifts to MZs belonging to a fixed step of the walk ($\phi_{m,n}=\phi_{n}, \forall m$), as in Figure 3b of the Supplementary Information.} {Experimental single- and two-photon output distributions are} summarized in Figure~\ref{Figura5}{a,b} for the boson{ic} and fermion{ic} 
case separately. The effect of a fully space-correlated dynamic randomness simulates the effect of an external classical environment. {In this case} one 
can show that the limiting distribution is a binomial centered in the
middle of the spatial axis and with width growing linearly with the square root of the number of steps. Thus the system undergoes a 
diffusion process in which the propagation  becomes equivalent to a purely classical random walk. The data of Figure~\ref{Figura5}{a-c} indeed confirm 
that in the case of dynamic disorder the walkers spread more easily on the BS tree {with respect to the static case, but spread less with respect to the ballistic diffusion of the ordered case}.

The last scenario we considered is the case of both space- and time-dependent disorder, {which we will call fluctuating disorder,} realized with random phase shifts {$\phi_{m,n}$} 
over the entire MZ network (see Fig. 3c of the Supplementary Information). In this configuration the diffusion process leads 
to a speckle pattern for the two-walkers wavefunction {(Figure \ref{Figura5}d-f). This case shows that the interaction with an external classical environment quenches the localization effect that would be induced by a lattice with static disorder.} 

Let us note that {the} experiments are performed 
on a single {phase map} realization of {each} disorder. Although there are still features that are linked to the particular choice of the (random{ly picked}) phase {maps}, the 
number of beam splitters is large enough to allow {the clear observation of the} differences between ballistic, diffusive and localized regimes.

We reported on the experimental realization of {a quantum simulator based on discrete quantum walks of entangled particles in integrated photonic circuits}. By properly 
engineering the phase shifts at the output ports of the BSs  and by changing the number {of QW steps,} we were able to follow in real time the evolution towards 
Anderson localization. The symmetry of the total wavefunction (Fermi- or Bose-like) clearly affects the localization properties. Fermi statistics helps localization {despite the fermions tendency to antibunching}. The quantum simulation we performed {will help} to 
ascertain the efficiency of quantum algorithms with entangled particles on realistic quantum walks. {The capability of our technology to implement arbitrary phase maps in QWs paves the way to the experimental quantum simulation of the quantum dynamics of multi-particle correlated systems and its ramifications towards the implementation of realistic universal quantum computation with quantum walks. }

This project was supported by FIRB- Futuro in Ricerca HYTEQ, PRIN 2009, IP-SOLID and ERC-Starting Grant: 3D-QUEST.

L.S, F.D.N., F.S., P.M., A.C., R.O. and R.R. conceived the
experimental approach for the simulation of the Anderson localization.
A.C., R.O., and R.R. fabricated the integrated devices. L.S., F.D.N.,
F.S., and P.M. carried out the experiments.
V.G. and R.F. contributed to the theoretical analysis on how statistics influence the localization. All the authors discussed the experimental
implementation and results and contributed to writing the paper.


\begin{thebibliography}{10}
\bibitem{footnote}
	These can be either in the form of  quantum entanglement of  the initial state   or in the form of the quantum statistics 
	of the involved particles.
\end{thebibliography}
\end{document}